\documentclass[prl,showpacs,superscriptaddress,twocolumn]{revtex4}

\usepackage{graphicx}

\begin{document}

\title{Tomographic Quantum Cryptography:\\
Equivalence of Quantum and Classical Key Distillation}

\author{Dagmar Bru\ss}
\affiliation{Institut f\"ur Theoretische Physik, Universit\"at Hannover, 
30167 Hannover, Germany}

\author{Matthias Christandl}
\affiliation{DAMTP, University of Cambridge, Cambridge~CB3~0WA, United Kingdom}

\author{Artur Ekert}
\affiliation{DAMTP, University of Cambridge, Cambridge~CB3~0WA, United Kingdom}
\affiliation{Department of Physics, National University of Singapore, 
Singapore 117\,542, Singapore}

\author{\mbox{Berthold-Georg~Englert}}
\affiliation{Department of Physics, National University of Singapore, 
Singapore 117\,542, Singapore}

\author{Dagomir Kaszlikowski}
\affiliation{Department of Physics, National University of Singapore, 
Singapore 117\,542, Singapore}

\author{Chiara Macchiavello}
\affiliation{Dipartimento di Fisica ``A.~Volta'', Universit\`a di  Pavia, 
27100 Pavia, Italy}

\date{31 March 2003}

\begin{abstract}
The security of a cryptographic key that is generated by communication 
through a noisy quantum channel relies on the ability to distill a
shorter secure key sequence from a longer insecure one. 
For an important class of protocols, which exploit  
tomographically complete measurements on entangled pairs of 
any dimension, we show that the noise threshold 
for classical advantage distillation is identical with the threshold for 
quantum entanglement distillation. 
As a consequence, the two distillation procedures are equivalent: 
neither offers a security advantage over the other.  
\end{abstract}


\maketitle

The ability to generate a secure cryptographic key, although the communication
employs a quantum channel with a high level of noise, is crucial for
all practical implementations of quantum cryptography.
To be on the safe side, one must assume that all noise results from
eavesdropping, that eavesdropper Eve has full knowledge of the cryptographic
protocol (the ``Kerckhoff principle'' of cryptology), 
and that she acquires as much
knowledge about the communication as is allowed by the laws of physics.
This leads immediately to the question of where is the noise threshold below
which a secure key can be generated at all.
We give a definite answer for an important class of protocols, restricting,
however, the discussion to incoherent attacks of the eavesdropper.

In the cryptographic protocol that we consider~\cite{Cherng}, 
Alice and Bob exploit entangled pairs of \emph{qunits}, 
that is: $n$-fold quantum alternatives, 
the case of $n=2$ being the elementary binary alternative of a qubit.
Alice measures on her qunit, and Bob on his, an observable randomly chosen
from their respective sets of $n+1$ observables that are tomographically
complete. 
Such sets surely exist for any dimension \cite{Wootters+1:89}.
Adopting the notation of \cite{MeanKing}, we write $\ket{m_k}$ for the $k$-th
eigenket of Alice's $m$-th observable and $\ket{\barr{m}_k}$ for the $k$-th
eigenket of Bob's $m$-th observable, 
whereby $m=0,1,\dots,n$ and $k=0,1,\dots,n-1$.

It is possible and expedient to choose these kets such that 
$\braket{0_j}{m_k}=\braket{\barr{m}_k}{\barr{0}_j}$ for all $m,j,k$, 
and then the maximally entangled
2-qunit state $\ket{\psi}$ that Alice and Bob wish to share,
\begin{equation}\label{eq:prot1}
\ket{\psi}= \frac{1}{\sqrt{n}}\sum_{k=0}^{n-1}\ket{0_k\barr{0}_k}
=\cdots= \frac{1}{\sqrt{n}}\sum_{k=0}^{n-1}\ket{n_k\barr{n}_k}\,,
\end{equation}
has the same appearance irrespective of the pair 
of observables that is used to define it.
Therefore, their measurement results in the matched bases 
(same value of $m$ for her and him) are perfectly correlated and can be used
for the generation of a key in an alphabet with $n$ letters.

On average, the measurement bases will be matched for a fraction $1/(n+1)$ of
the qunit pairs, and these data will supply the raw key sequence.
Alice and Bob use part of it together with all the other measurement data, 
acquired for mismatched bases, to perform quantum tomography on the two-qunit
state they are actually receiving from the source.
The tomographic completeness of the two sets of observables 
is crucial for this part of the procedure.

Alice and Bob assume that Eve distributes the qunits.
They accept the raw key only if the result of their state tomography 
is consistent with an admixture of the chaotic state to 
$\ket{\psi}\bra{\psi}$, thereby forcing Eve to use a symmetric strategy. 
In other words, they 
accept only a 2-qunit state $\rho$ of the form
\begin{equation}
  \label{eq:prot2}
  \rho=(\beta_0-\beta_1)\ket{\psi}\bra{\psi}+\frac{\beta_1}{n}I\,,
\quad \beta_0+(n-1)\beta_1=1\,,
\end{equation}
where $I$ is the 2-qunit identity operator, 
$\beta_0$ is the probability that Bob gets the same value as Alice when
the bases match, and $\beta_1$ is the probability that he gets a particular
other one. 
Since $\beta_0=\beta_1=1/n$ when there are no correlations whatsoever
between their measurement results, we take $\beta_0>1/n>\beta_1$ for granted.

Although Eve fully controls the 2-qunit source, 
she is not free in her actions, because the state 
received by Alice and Bob must be of the form (\ref{eq:prot2}).
One finds \cite{Cherng} that, therefore, the best Eve can do is to prepare 
an entangled pure state of the form 
\begin{equation}\ket{\Psi}=
\sqrt{\frac{\beta_0}{n}}
\sum_{k=0}^{n-1}\ket{0_k\barr{0}_k}\ket{\mathrm{E}_{kk}}+
\sqrt{\frac{\beta_1}{n}}
\sum_{k\neq l}\ket{0_k\barr{0}_l}\ket{\mathrm{E}_{kl}}\,,  
\label{eq:prot3}
\end{equation}
where her normalized ancilla states $\ket{\mathrm{E}_{kl}}$ are such
that those with $k\neq l$ are orthogonal to all others, whereas those 
with $k=l$ are not orthogonal among themselves, but obey 
$\braket{\mathrm{E}_{kk}}{\mathrm{E}_{ll}}%
=1-(\beta_1/\beta_0)(1-\delta_{kl})$.
Thus the summations in (\ref{eq:prot3}) constitute two orthogonal components of
$\ket{\Psi}$. 
The $n$-dimensional first component is relevant for
establishing the cryptographic key, the $n(n-1)$-di\-men\-sio\-nal second 
component is just noise to Alice and Bob.  

We note that the invariance of $\ket{\psi}$ under bases permutations 
is also possessed by $\ket{\Psi}$.
Rather than referring to the $0$-th pair of observables, we could just as 
well use the joint eigenkets $\ket{m_k\barr{m}_l}$ of any other pair 
in conjunction with a suitable unitary redefinition of the ancilla states.  

After Alice and Bob have given public notice of 
the observables they measured for each qunit pair, it is Eve's task to infer
their measurement results---their nit values---whenever the bases match. 
To this end she must be able to identify her ancilla states. 
(Remember that we are only considering incoherent eavesdropping attacks.)
Owing to the structure of $\ket{\Psi}$ she can distinguish unambiguously 
all the states belonging to the second orthogonal component,
and so she can correctly infer Alice's and Bob's nit values 
if they are different.
But if they are the same, Eve has to distinguish the $\ket{\mathrm{E}_{kk}}$  
of the first component, and then she cannot avoid errors because these states
are not orthogonal to each other.  
In this situation, she minimizes her error probability by performing 
the so called \emph{square-root measurement} \cite{Chefles:00}.

We denote by $\eta_0$ and $\eta_1=(1-\eta_0)/(n-1)$ the probabilities
that Eve infers the nit value correctly or gets a particular wrong one,
respectively, provided that Bob's nit value is the same as Alice's.
They  are related to Bob's probabilities $\beta_0$ and $\beta_1$ 
of (\ref{eq:prot2}) by
\begin{equation}\label{eq:prot4}
\sqrt{\eta_0}-\sqrt{\eta_1}=\sqrt{\beta_1/\beta_0}\,.
\end{equation}
Note that this expresses a certain complementarity between Bob's and Eve's
respective knowledge about Alice's nit values.
If Bob's values agree perfectly with Alice's ($\beta_0=1$, $\beta_1=0$), then
Eve's values are completely random ($\eta_0=\eta_1=1/n$),
and conversely $\eta_0=1$, $\eta_1=0$ implies $\beta_0=\beta_1=1/n$.
In the more interesting intermediate situations we have $\eta_0>1/n>\eta_1$.

For single-particle protocols with qubits ($n=2$) or qutrits ($n=3$), 
the relation (\ref{eq:prot4}) is well established \cite{Bruss+1:02,notation}, 
and has been conjectured to hold for arbitrary dimensions 
\cite{Bruss+1:02,Acin+2:03}. 
This conjecture is proved in \cite{Cherng}.

According to the Csisz\'ar--K\"orner (CK) Theorem \cite{Csiszar+1:78}, a secure
key sequence can be extracted from the raw key sequence if the mutual
information between Alice and Bob exceeds the mutual information between
either one of them and Eve.
This requires that Bob's and Eve's probabilities are such that
\begin{eqnarray}
  \label{eq:CKth}
  \nu&\equiv&\beta_0\log_n\beta_0+(1-\beta_0)\log_n\beta_1
\nonumber\\&&
-\beta_0\bigl[\eta_0\log_n\eta_0+(1-\eta_0)\log_n\eta_1\bigr]>0\,,
\end{eqnarray}
and then $\nu$ is the yield of the CK procedure, the fraction of nit values
that make it from the raw key sequence to the secure one.
Since (\ref{eq:prot4}) implies that $\eta_0,\eta_1\to1/n$ as $\beta_0\to1$,
this condition is surely met if $\beta_0$ is sufficiently large.
If, however, there is too much noise in the 2-qunit state (\ref{eq:prot2}),
the CK theorem is not immediately applicable.
Rather, Alice and Bob must select a subsequence of nit values in a systematic
way such that the CK theorem applies to the resulting ``distilled key.''  

One method at their disposal for this purpose 
is \emph{entanglement distillation} (ED), a quantum procedure by which  
they produce a smaller number of qunit pairs with stronger entanglement, by
means of local operations and classical communication \cite{Deutsch+5:96}. 
Thus they can reach a $\beta_0$ value for which the CK theorem is applicable,
before they measure their respective observables. 
For states of the particularly simple structure (\ref{eq:prot2}), 
ED will be successful if
\begin{equation}
  \label{eq:prot5}
  \beta_0>2\beta_1
\end{equation}
and only then \cite{Horodeccy}.
If Eve can perfectly compensate for the back effect of ED on the ancillas, 
relation (\ref{eq:prot4}) also applies after ED.
If she cannot, it turns into an inequality, tersely: $=\,\to\,<$.
 
Alternatively, Alice and Bob can produce their raw key sequences without
any subensemble selection, 
and then perform \emph{advantage distillation} (AD), 
a procedure of classical (i.e.\ non-quantum) cryptography \cite{Maurer+al}. 
As we shall see below, for $\beta_0,\eta_0$ values that obey (\ref{eq:prot4}),
AD is successful whenever (\ref{eq:prot5}) holds, and
only then, so that the thresholds for ED and AD are the same  
\cite{Gisin+1:99}.
As a consequence of this coincidence, ED and AD are equivalent in the sense
that neither offers a security advantage over the other. 

Both ED and AD require classical two-way communication, but once
the CK theorem becomes applicable, one-way communication suffices.
We leave it as a moot point which method makes better use of the resources
because the standard versions of both are very wasteful and hardly suited
for practical implementation \cite{efficiency}.    

The AD protocol is as follows.
Alice and Bob divide their raw strings of nit values into blocks 
of length $L$.
For each block, Alice casts a $n$-sided die and then
adds, modulo $n$, the random value thus found to the given block.
Then she sends these modified blocks to Bob through
a public, but authenticated channel. 
Bob subtracts, modulo $n$, his corresponding blocks.
Whenever he obtains a block consisting of $L$ identical nit values,
he enters this value into his distilled sequence.
If, however, different values appear in a block, he disregards it.
He tells Alice, through the public channel, which blocks contribute to the
distilled key and which don't.
She in turn then forms her own distilled sequence from the random values that
she added to the blocks that Bob did not discard.

The two distilled sequences are identical, except at the rare positions,
where Bob's whole block consisted of $L$ wrong nit values of the same kind.
Since there are $n-1$ different wrong nit values, 
the relative frequency with which a particular one occurs 
in the distilled sequence is
\begin{equation}
  \label{eq:clAD0}
  B_L=\frac{\beta_1^L}{\beta_0^L+(n-1)\beta_1^L}\,,
\end{equation}
which is, so to say, the new value of $\beta_1$ after AD. 
Since $\beta_1<\beta_0$, $B_L$ decreases exponentially with the
block length $L$,
\begin{equation}
  \label{eq:clAD0'}
  \lim_{L\to\infty}\frac{B_{L+1}}{B_L}=\frac{\beta_1}{\beta_0}\,.
\end{equation}

Whenever Alice and Bob end up with a pair of different nit values after AD,
Eve knows both values correctly, because she knows the values for each nit
pair of the two blocks in question.
But when Alice and Bob get the same value, 
which is the much more frequent situation for long blocks, 
Eve cannot be completely sure about any nit value in the blocks.
Her best strategy is then to subtract Alice's block from her 
corresponding block, that is, to do what Bob does. 
Typically, Eve's block is inhomogeneous after the subtraction, 
and it is highly likely that the correct nit value is the value 
that occurs more often than any other. 
So, she decides by a majority vote which nit value to assign: 
She bets on the value that appears most frequently in the block, and if
there are several most frequent values, she picks one of them at random.

To find her probability for assigning the right value, we first note that a
block with $m$ correct values and $k_1$, $k_2$, \dots, $k_{n-1}$ 
ones of the $n-1$ wrong kinds, respectively, occurs with a relative frequency 
that is given by
\begin{eqnarray}
  \label{eq:clAD1}
&&  \frac{L!\ \eta_0^{m}\eta_1^{L-m}}{m!k_1!\cdots k_{n-1}!}
\delta_{L-m,k_1+k_2+\cdots+k_{n-1}}
\nonumber\\
&=&\binom{L}{m}\eta_0^{m}
\left.\left(\frac{\partial}{\partial x}\right)^{L-m}\;
\prod_{j=1}^{n-1}\frac{(\eta_1x)^{k_j}}{k_j!}\right|_{x=0}\,,
\end{eqnarray}
where the Kronecker delta symbol enforces the constraint 
$L-m=k_1+k_2+\cdots+k_{n-1}$. 

Second, we note that Eve surely assigns a wrong value whenever 
$m<\max\{k_1,\ldots,k_{n-1}\}$, and in the situation of 
$m\geq\max\{k_1,\ldots,k_{n-1}\}$ she assigns the right value 
with probability $1/(l+1)$ where $l$ is the count
of $k_j$'s  that are equal to $m$. 
With the combinatorial factor
\begin{equation}
  \label{eq:clAD3}
  \binom{n-1}{l}=\binom{n}{l+1}\frac{l+1}{n}
\end{equation}
taken into account, the summation over $m$ and all $k_j$ thus gives
\begin{eqnarray}
  \label{eq:clAD4}
&& 1-(n-1)E_L=\sum_{m=0}^{L}\binom{L}{m}\eta_0^{m}
\left(\frac{\partial}{\partial x}\right)^{L-m}
\nonumber\\&&\hspace*{2em}\times
\frac{1}{n}\sum_{l=0}^{n-1}\binom{n}{l+1}
\left[\frac{(\eta_1x)^{m}}{m!}\right]^l
\left[\sum_{k=0}^{m-1}\frac{(\eta_1x)^{k}}{k!}\right]^{n-1-l}
_{\Bigr|_{\mbox{\scriptsize$x=0$}}}
\nonumber\\
\end{eqnarray}
for the probability that Eve assigns the right nit value to a string of
length $L$, and $E_L$ is then the probability that she gets 
a particular one of the $n-1$ wrong values.
Parroting the remark after (\ref{eq:clAD0}), we note that $E_L$ is, so to say,
the new value of $\eta_1$ after AD.

We use the generating function
\begin{equation}
  \label{eq:clAD5}
  E(t)\equiv\sum_{L=0}^\infty\frac{t^L}{L!}E_L
\end{equation}
to deal with these probabilities as a set.
It is given by
\begin{eqnarray}
  \label{eq:clAD6}
  E(t)&=&\frac{e^t}{n-1}\sum_{m=0}^\infty\frac{(\eta_0t)^m}{m!}e^{-\eta_0t}
\nonumber\\&&\times
\left(1-\frac{\bigl[w_m(\eta_1t)\bigr]^n-\bigl[w_{m-1}(\eta_1t)\bigr]^n}
{n\bigl[w_m(\eta_1t)-w_{m-1}(\eta_1t)\bigr]}
\right)
\end{eqnarray}
where
\begin{equation}
  \label{eq:clAD7}
  w_m(x)=\sum_{k=0}^m\frac{x^k}{k!}e^{-x}
\end{equation}
is the partially summed Poisson distribution.

For the comparison with (\ref{eq:clAD0'}),
the quantity of primary interest is the limit
\begin{equation}
  \label{eq:clAD8}
  \lim_{L\to\infty}\frac{E_{L+1}}{E_L}
=\lim_{t\to\infty}\frac{\partial}{\partial t}\log E(t)\,,
\end{equation}
which directs our attention to the large-$t$ behavior of $E(t)$.
Now, for large $t$ the Poisson distribution in $m$, by which the
parenthesized difference is weighted in (\ref{eq:clAD6}), has its peak at 
$m\simeq\eta_0t>\eta_1t$, and the relative width of this peak shrinks with
growing $t$. 
Accordingly, all relevant contributions to the sum in (\ref{eq:clAD6}) have
$m>\eta_1t$, so that the approximation 
\begin{equation}
  \label{eq:clAD7'}
  w_m(\eta_1t)=w_{m-1}(\eta_1t)+\frac{(\eta_1t)^m}{m!}e^{-\eta_1t}\simeq1 
\end{equation}
is permissible, and
\begin{equation}
  \label{eq:clAD9}
   E(t)\simeq\frac{1}{2}e^{(1-\eta_0-\eta_1)t}
      \,\mathrm{I}_0(2\sqrt{\eta_0\eta_1}\,t)\quad\text{for $t\gg1$}
\end{equation}
obtains.
Since there is no difference between the modified Bessel function
$\mathrm{I}_0(z)$ and its derivative $\mathrm{I}_1(z)$ when $z\gg1$, 
this tells us that
\begin{equation}
  \label{eq:clAD10}
  \lim_{L\to\infty}\frac{E_{L+1}}{E_L}
=1-\bigl(\sqrt{\eta_0}-\sqrt{\eta_1}\,\bigr)^2\,.
\end{equation}

\begin{figure}[!t]
\vspace*{1ex}
\centerline{\includegraphics{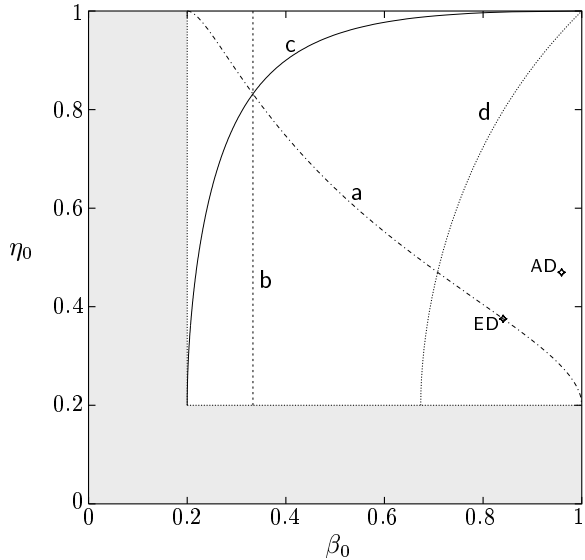}}
\caption{\label{fig:3coinc}%
The three-fold coincidence, for $n=5$, in a plot of Bob's probability 
$\beta_0$ vs.\ Eve's probability $\eta_0$.
The relevant values of $\beta_0>1/n$ and $\eta_0>1/n$ are outside 
the gray area.
The dash-dotted curve~\textsf{a} identifies the ${\beta_0,\eta_0}$ pairs for
which (\ref{eq:prot4}) holds.
To the right of the dashed vertical line~\textsf{b}, condition
(\ref{eq:prot5}) is obeyed and ED is possible.
By contrast, AD 
can be successfully performed below the solid-line curve~\textsf{c} 
which marks the border stated in (\ref{eq:clAD11}).
All three lines intersect at $\beta_0=1/3=0.33$,
$\eta_0=(11+4\sqrt{6})/25=0.83$,
so that the part of curve~\textsf{a} that is to the right of curve~\textsf{b}
is also the part that is below curve~\textsf{c}.
--- 
The CK theorem is applicable to ${\beta_0,\eta_0}$ values
below the dotted curve~\textsf{d} that results from (\ref{eq:CKth}).
It intersects curve~\textsf{a} at $\beta_0=0.708$, $\eta_0=0.470$.
From there, a single ED step takes one to the $\diamond$ point on
curve~\textsf{a}, whereas AD with $L=2$ moves one horizontally 
to the $\diamond$ point further right (see \cite{efficiency}).
}
\end{figure}

In conjunction with (\ref{eq:clAD0'}), it follows that 
AD of this kind will be successful if
\begin{equation}
  \label{eq:clAD11}
 \frac{\beta_1}{\beta_0}<1-\bigl(\sqrt{\eta_0}-\sqrt{\eta_1}\,\bigr)^2
\end{equation}
holds because then Bob's error probability gets exponentially smaller 
than Eve's with increasing block length $L$, and the distilled key sequence
will meet the requirements of the CK theorem if $L$ is chosen large enough.
Now, if (\ref{eq:prot4}) relates Eve's probabilities to Bob's, 
as it is the case for the probabilities originating in Eve's 
source state (\ref{eq:prot3}), the
threshold condition (\ref{eq:clAD11}) for classical AD is, indeed, 
identical with the threshold condition (\ref{eq:prot5}) 
for quantum ED, as we have asserted above.
This remarkable coincidence is illustrated in Fig.~\ref{fig:3coinc}.

It is important to note that, despite its simplicity and its lack of
efficiency, the AD scheme considered correctly identifies the threshold point
on curve~\textsf{a} in Fig.~\ref{fig:3coinc}.  
For, if $(\beta_0,\eta_0)$ is to the left of the triple-coincidence point, 
the 2-qunit state (\ref{eq:prot2}) is separable.
Eve can then blend it from product states and can so ensure that there is 
no useful mutual information between Alice and Bob,
and without it they cannot generate a secure key.

For $n=3$, 
a more involved argument about the same matter is given in \cite{Acin+2:03}. 
In fact, while our paper was being written, we became aware of \cite{Acin+2:03}
where some of our results are conjectured and identical conclusions are 
reached.
One wonders, of course, whether the surprising equivalence between 
classical and quantum distillation is more than just the coincidence 
as which it appears here and in \cite{Acin+2:03}. 
Perhaps it hints at a deeper connection between these fundamentally 
different procedures.

Finally, one might wonder if Eve has a better procedure at her disposal 
than the square-root measurement that is the basis of our analysis.
For the following reasons we think she does not. 
The error-minimizing strategy takes full advantage of the built-in symmetry 
of the tomographic protocol. 
For all other fully symmetric eavesdropping attacks, the CK region is reached 
at smaller $\beta_0$ values, and successful AD is possible for all of them if
the ED threshold (\ref{eq:prot5}) is crossed \cite{symmetric}.
It seems, therefore, rather reasonable that the error-minimizing strategy is
Eve's optimal choice.

\begin{acknowledgments}
We are indebted to Antonio Ac\'\i{}n, Thomas Durt, Leong Chuan Kwek, 
and Norbert L\"utkenhaus  for sharing their insights with us.
DB, MC, and CM gratefully acknowledge the splendid 
hospitality 
extended to them at the National University of Singapore. 
This work was supported by A$^*$Star Grant No.\ 012-104-0040.
\end{acknowledgments}


\begin{thebibliography}{99}
\bibitem{efficiency}
A bit of evidence---not conclusive, of course---that AD could be more
efficient than ED is provided by the example of taking
$n=5$ and choosing $\beta_0,\eta_0$ such that $\nu=0$ in (\ref{eq:CKth}). 
Then, accounting for both the efficiency  of the distillation step and the CK
efficiency (\ref{eq:CKth}) of the distillation product,
a single ED step of the simple two-pair kind discussed in \cite{Horodeccy} 
gives a total yield of $\frac{1}{2}\times0.522\times0.388=10.1\%$, 
whereas a single AD step with $L=2$ has a yield of 
$\frac{1}{2}\times0.522\times0.709=18.5\%$.
\bibitem{symmetric}
This can be demonstrated by a suitable generalization of our analysis 
of AD for the error-minimizing strategy.
\end{thebibliography}
\end{document}